# Early warning signs in social-ecological networks

*Samir Suweis[1] and Paolo D'Odorico[2]*

[1] Department of Physics and Astronomy, University of Padova. suweis@pd.infn.it

[2] Department of Environmental Science, University of Virginia. paolo@virginia.edu

**Social ecological systems are often difficult to investigate and manage because of their inherent complexity[1]. Small variations in external drivers can lead to abrupt changes associated with instabilities and bifurcations in the underlying dynamics[2-4]. Anticipating critical transitions and divergence from the present state of the system is particularly crucial to the prevention or mitigation of the effects of unwanted and irreversible changes[5-10]. Recent research in ecology has focused on leading indicators of regime shift in ecosystems characterized by one state variable[5,7,11,12]. The case of systems with several mutually interacting components, however, has remained poorly investigated[13], while the connection between network stability and research on indicators for loss of resilience has been elusive[14]. Here we develop a theoretical framework to analyze early warning signs of instability and regime shift in social ecological networks. We provide analytical expressions for a set of precursors of instability in social ecological systems with additive noise for a variety of network structures. In particular, we show that the covariance matrix of the dynamics can effectively anticipate the emergence of instability. We also compare signals of early warning based on the dynamics of suitably selected nodes, to indicators based on the integrated behavior of the whole network. We find that the performances of these indicators are affected by the network structure and the type of interaction among nodes. These results provide new advances in multidimensional early warning analysis and offer a framework to evaluate the resilience of social ecological networks.**

We consider a social-ecological system with *N* components (nodes) coupled through a set of links. The state of the system is expressed by the vector **x** of length *N*, whose terms $x_i$ represent the state of node *i*. The local stability of a state $\mathbf{x}^*$ is evaluated through a linearization, $\frac{dy}{dt} = \boldsymbol{A} \cdot \boldsymbol{y}$, where **y**=**x**−**x*** is the displacement of **x** from **x***; **A** is the $N \times N$ matrix expressing the interactions among nodes in the (linearized) dynamics (see Methods). In population ecology this framework is typically used to express the dynamics of a community of $N$ populations interacting according to the relationships determined by the matrix **A**, often known as "*community matrix*"[2,15,16]; likewise, in social systems **A** describes the network of interactions (e.g., trade, migration, flow of information among people, groups of individuals, or countries[17-19]). The off-diagonal terms of **A** determine the pairs of interacting nodes as well as the strength of their interaction. The dynamics are stable if the maximum eigenvalue, $\lambda_{max}$, of **A** is negative.

Classic ecological theories[2,3] have considered the case of networks with randomly connected nodes (with a certain probability, $C$). The strength ($p$) of the interactions between them is represented by a zero-mean random variable of variance $\sigma^2$. May[2,3] showed that random networks become unstable as connectivity (i.e., $C$), size (i.e., $N$) or strength variance increase. These findings were recently generalized to the case of networks with prescribed structures (e.g., predator-prey, competitive or mutualistic interactions)[15]: the stability/instability of the system was found to strongly depend on the network structure as well as on connectivity, strength variance, and system size.

More in general, the off-diagonal terms of **A** may result from a set of "rules" expressed as a function of a few parameters of which connectivity and strength variance are just an example. Changes in the structure and intensity of the interactions correspond to variations in these parameters, which, in turn, can lead to instability by modifying the community matrix and its eigenvalues. How can we evaluate whether ongoing changes in the interactions within a social-ecological network are reducing its resilience? Is there a way to use measurable quantities to determine whether the system is about to become unstable?

In one-dimensional systems leading indicators are typically associated with behaviors resulting from the eigenvalue tending to zero at the onset of instability. This effect entails a slower return to equilibrium after a "small" perturbation[11,20]. Known as "critical slowing down", this phenomenon exists also in systems with multiple interacting components, though it is hard to recognize and therefore it does not constitute an effective leading indicator of instability. In fact, in "real world" applications the equations driving the dynamics are not known and, therefore, the network nodes in which slowing down is expected to occur are not known a priory. Critical slowing down, however, has been related to an increase in variance and autocorrelation in the state variable of one dimensional systems[5,7,21]. Here we provide a theoretical framework to investigate early-warnings in the variance, autocorrelation, and power spectrum of multi-dimensional systems with interactions described by a given network structure.

We generate networks of size $N$, with a variety of architectures for **A** (see Methods), and reach instability either by keeping constant the connectivity, $C$, while changing the strength of the interactions, $p$, or by varying $C$ for a fixed $p$[2,15]. We then determine the analytical relationship between the steady state covariance matrix, $\mathbf{S_y}$, of **y** and the eigenvalues of the matrix **A**. Similarly, we express the time-lag correlation, $\mathbf{\rho_y}$, and the power spectrum, $\mathbf{P_y}$, of **y** as function of **A** and its eigenvalues. We find that the elements of both $\mathbf{S_y}$ and $\mathbf{\rho_y}$ increase as the system approaches instability (i.e., $\lambda_{max} \to 0$). Therefore, we investigate potential indicators for early warning in the behavior of suitable components of $\mathbf{S_y}$, $\mathbf{\rho_y}$ and $\mathbf{P_y}$ for $\lambda_{max} \to 0$. To that end we first consider the components of $\mathbf{S_y}$ corresponding to the most connected, the most central[22] and the least connected nodes of the network. We also consider indicators based on the properties of the entire network, such as the maximum and the difference between the maximum and minimum of the matrix $\mathbf{S_y}$.

We find that most of the indicators based on the covariance matrix, $S_y$, have a non-trivial dependence on $\lambda_{max}$ (see Figure 1). The maximum element of $S_y$ ($Max[S_y]$) provides the most effective indicator of early warning in most networks, except for the case of random networks, in which $Max[S_y] - Min[S_y]$ exhibits a stronger increase at the onset of instability (Figure 1). In mutualistic (++) networks $Max[S_y]$ corresponds to the most connected node (the "hub"), regardless of their topological structure (Supplementary Information). All these indicators based on $S_y$ improve their performances when the size, $N$, of the network increases, as shown by the comparison between main panels and insets in Figure 1 (see also Supplementary Information). Thus our ability to detect early warning signs and predict tipping points is enhanced in more diverse systems[14].

We also look at the relationship between the maximum element of the time-lag correlation matrix, $\rho_y(\Delta)$ (where $\Delta$ is the time lag), and the maximum eigenvalue, $\lambda_{max}$, for different values of $\Delta$, $p$ and $C$. We find that, although significant, these indicators are less efficient with respect to the case with zero time-lag (i.e., indicators based on $S_y$). Finally, the power spectrum does not appear to be an effective indicator, as we identified only weak changes in $P_y$ for increasing values of $p$ and $\lambda_{max}$ (see Supplementary Information). Therefore, here we focus on early warning signs provided by the way $Max[S_y]$ varies as a function of changes in $\lambda_{max}$.

A warning sign is effective if (1) it appears in time to prevent (or prepare for) the occurrence of instability[23,24]; (2) it relies on a well-defined and easy to recognize indicator (e.g., a detectable or significant increase in variance[23,25]); and (3) it does not give false positives (or false negatives)[26]. We use these criteria to evaluate the effectiveness of $Max[S_y]$ as a leading indicator of instability with different network structures and levels of noise[24].

To investigate the effect of noise, we first consider the "mean-field" case of networks in which the absolute value of the interaction strength between connected nodes is a constant, $p$; we gradually increase $p$ or $C$ until the real part of the maximum eigenvalue of $A$ becomes positive[15]. We observe (Fig. 2) a consistent increase in $Max[S_y]$ for all network structures, regardless of whether instability is attained by increasing interaction strength or connectivity. The network structure, however, affects the timeliness of $Max[S_y]$ as a leading indicator. In fact, $Max[S_y]$ exhibits a more defined increase and a better anticipation of the onset of instability in the case of random networks than with all the other structures. In the case of these "mean field" networks we did not consider the antagonistic structure because antagonistic networks with constant interaction strength (in absolute value) are always stable regardless of the parameters $p$ and $C$ (see Supplementary Information).

Likewise, in the case of random interaction strengths (see Supplementary Information) $Max[S_y]$ exhibits a well-defined increase and a better anticipation of the instability in random networks than with the more organized structures typical of ecological or social systems (Figure 3). The seemingly weaker increase in $Max[S_y]$ observed in the social ecological networks is only an apparent effect of the scale. Indeed, as it will be shown later, suitable detection criteria of early warnings are more successful in mutualistic networks than in their random counterparts. It is also observed that noise has the effect of amplifying the intensity of the warning sign (compare the scales in Figs. 2 and 3), while inducing some weak random fluctuations with no substantial impact on the overall behavior of $Max[S_y]$ at the onset of instability (see Supplementary Information). In scale free networks the increase in $Max[S_y]$ (Fig. 3) is again only apparently muted. In fact, in these networks detection criteria are quite successful in recognizing early warning signs (Figure 4); moreover, local indicators (e.g., the variance of the most central node) can exhibit a more pronounced increase that can be used as an early warning sign of instability (Figure 1 and Supplementary Informations).

Overall, the performances of $Max[S_y]$ as a leading indicator of instability change between random, antagonistic, mutualistic/social networks. This indicator gives an earlier and "sharper" warning sign in random than mutualistic and social networks. The warning sign, however, is harder to detect and is more likely to be missed in random and antagonistic networks than in their mutualistic or social counterparts. Thus, by affecting the probability that early warnings are missed, the sign of the interactions within the network determines the consistency and reliability of this leading indicator. In fact, different realizations of the same network dynamics can yield different results in the behavior of $Max[S_y]$ and thus this indicator might not detect in useful advance the emergence of instability. The probability of true positives is close to 100% (i.e., negligible probability of false negatives) in mutualistic networks, and much smaller in random and antagonistic (predator-prey, cascade or compartment) networks (Figure 4). Thus, while mutualistic networks are less stable than their antagonistic counterparts[15], their instability can be predicted with less uncertainty. An increase in $Max[S_y]$, however, would not provide information on how close the system is to the onset of instability. Rather, it would just indicate that the system is losing resilience and approaching unstable conditions[23]. Therefore, in contrast to previous expectations[14], it is not the heterogeneity in the topology of the network that plays a key role in the abruptness of critical transitions and our ability to predict them. Rather, it is the type of interactions between the nodes that determines how networks respond to external perturbations. In fact, there is a trade-off between local and systemic resilience: mutualism (++) is associated with a reduced local stability and resilience of the system[15], but does not induce abrupt critical transitions. In contrast, networks with mixtures of interaction types (+-,++--) exhibit shorter recovery times after displacement from equilibrium (i.e., a stronger local resilience), but in these systems the emergence of systemic instability and critical transitions is more difficult to predict in useful advance.

This study combines stability theories from community ecology[2,15] to recent research on indicators of critical transition[7,9,14], and develops a unified framework that offers a new perspective method for the evaluation of the resilience and anticipation of instability in social ecological networks.

## Methods Summary

**Early Warning in Complex Networks.** We consider a network with $N$ interacting nodes. The state of the system, $\mathbf{x}=\{x_1, x_2, \ldots x_N\}$, is governed by the set of coupled dynamical equations with additive noise: $d\mathbf{x} = \mathbf{f}(\mathbf{x},p,C)dt + v\,\mathbf{I}\,dW$, where $\mathbf{f}=\{f_1, f_2, \ldots, f_N\}$ is a $N$-dimensional vector function expressing the deterministic component of the dynamics of $\mathbf{x}$, as a function of a set of parameters, $p$ and $C$; $\mathbf{I}$ is the identity matrix, and $v\,dW$ is the stochastic driver represented by a white Gaussian noise of mean zero and intensity $v dt$. This framework can also be generalized to the case in which $\mathbf{I}$ is replaced by a non-diagonal matrix (i.e., with correlation among the noise terms driving the dynamics of each node).

If we consider a small perturbation $\mathbf{y}$ forcing the system away from its equilibrium point $\mathbf{x}^*$ (i.e., $\mathbf{y}=\mathbf{x}-\mathbf{x}^*$), inserting $\mathbf{x}=\mathbf{x}^*+\mathbf{y}$ in the above equation and linearizing $\mathbf{f}(\mathbf{x}^*+\mathbf{y}, p, C)$ around $\mathbf{x}^*$ we obtain

$$d\mathbf{y} = \mathbf{A}(p)\mathbf{y}dt + \mathbf{I}v\,dW, \qquad (1)$$

where $A_{ij} = [\partial f_i / \partial x_j]_{x=x^*}$. Eq. (1) is a multivariate Ornstein–Uhlenbeck process[27]. Following Allesina[15], we build $\mathbf{A}$ for eleven different architectures (see Supplementary Information)

The stable states, $\mathbf{x}^*$, of stochastic dynamics driven by additive noise are the same as those of their deterministic counterparts, $\frac{d\mathbf{x}}{dt} = \mathbf{f}(\mathbf{x},p,C)$ [28]. These states are stable if the maximum real part of the eigenvalues of $\mathbf{A}$ is negative. To identify early warning signs of network instability, we relate the steady state covariance matrix $\mathbf{S}_\mathbf{y} = \langle \mathbf{y}_s, \mathbf{y}_s^T \rangle$ to the eigenvalues, $\lambda$ of $\mathbf{A}$, where $\mathbf{y}_s$ is calculated from the steady state solution of Eq. (1). The $(i, j)$ element of $\mathbf{S}_\mathbf{y}$ is:

$S_y(i,j) = \langle y_i y_j \rangle - \langle y_i \rangle \langle y_j \rangle$, where $\langle \rangle$ represents the average. The covariance matrix of the stationary dynamics of the system can be obtained as the solution of the equation[27]:

$$A(p,C)S_y + S_y A^T(p,C) = -\nu I \qquad (2)$$

$S_y$ is a function of the linearization matrix, $A(p,C)$, which, in turn, depends on the control parameters ($p$ or $C$). At the onset of instability (i.e., as $Re(\lambda_{max}) \to 0$) the maximum element of the covariance matrix, $S_y$, of $y$ increases (see Supplementary Information). More details on the time-lag correlation, power spectrum, and the generation of social-ecological networks can be found in the Supplementary Information.

**Detection of early warning** To evaluate whether the onset to instability can be anticipated in time by an increase in $Max[S_y]$ (or in other suitably chosen elements of $S_y$), we test the correlation[25] between $Max[S_y]$ and the control parameter ($p$ or $C$) that is gradually varied to increase $Max[Re(\lambda)]$ up to a given threshold (here chosen equal to $-0.2$). If the correlation (evaluated with the Kendall-$\tau$ test) is significant and greater than 0.5, the increase in $Max[S_y]$ is interpreted as an early warning sign. We repeat this analysis for 1000 realizations of the random interaction strength network and determine the distribution of correlations along with the number of realizations with positive warning sign.

29. Albert R. & Barabasi A-L. Statistical mechanics of complex network. *Rev Mod Phys* **74**(1):47 (2002).

## 30. Figures

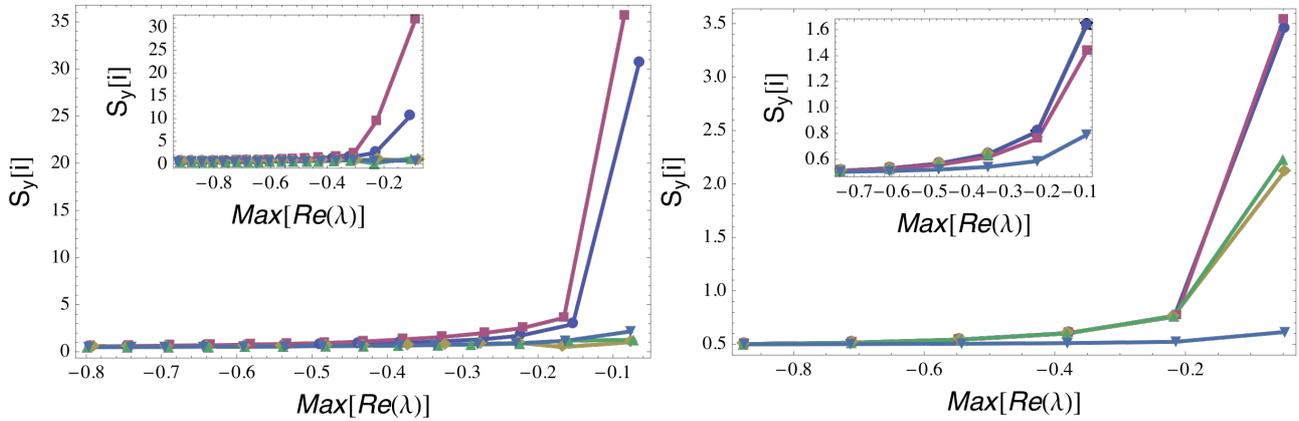

**Figure 1.** Leading indicators of instability based on different elements of the covariance matrix ($S_y$), including the maximum (in absolute value) element, $Max[S_y]$ (purple), the difference between $Max[S_y]$ and $Min[S_y]$ (pink), the element of $S_y$ corresponding to the most connected (gold), least connected (blue) or highest eigenvector centrality[22] (green) network node. Random (*left*) and scale free (right)[29] network generated with $N=50$ and $C=0.1$ (main panels) and $N=10$ and $C=0.5$ (insets). Instability (i.e., decrease in $Max[Re(\lambda)]$) is attained by increasing the interaction strength *p* (mean field case). The figures represent average behavior over 100 realizations.

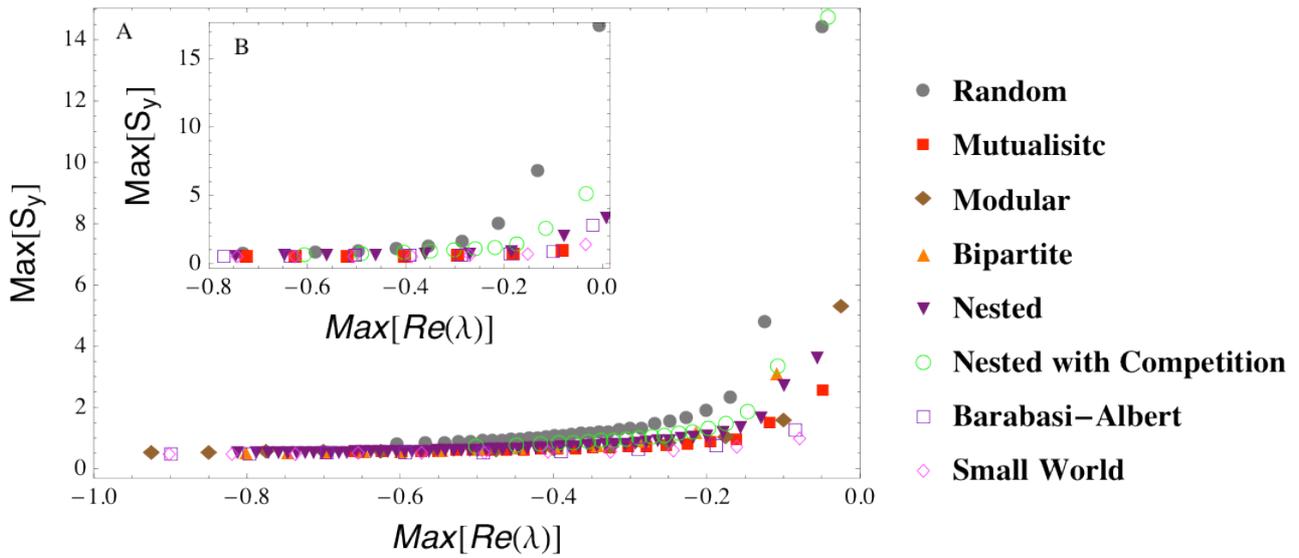

**Figure 2.** *Max*[**S$_y$**] as a leading indicator of instability in a "mean field" network with constant interaction intensity (in absolute value), *p*. Instability is attained by increasing *p* (main panel A, with *N*=20, *C*=0.2) or *C* (inset B, with *N*=20, and *C* increasing from 0.1 to 1) with different network structures. The figures represent average behavior over 1000 realizations.

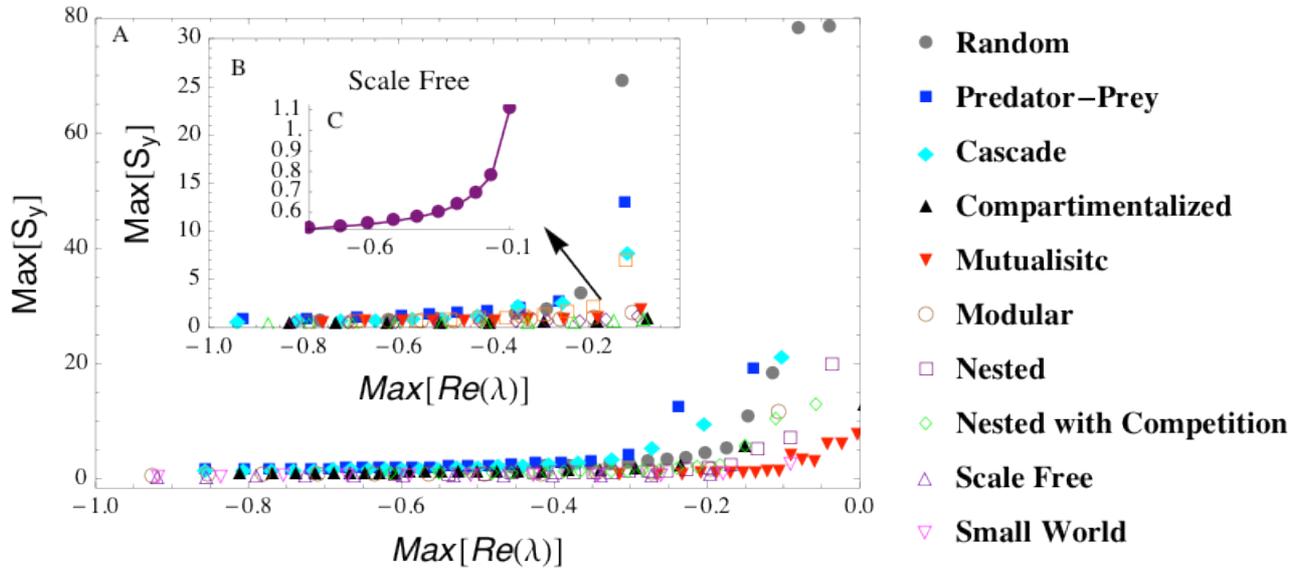

**Figure 3.** A) Case with random interaction strength (see methods). Main panel: instability is reached by increasing *p* (with *N*=20; *C*=0.2). First inset (B): *p* is constant while *C* increases between 0.1 and 1. C) Same as the first inset (B) but only for the scale-free network (notice the different scale on the vertical axis). The figures represent average behavior over 1000 realizations.

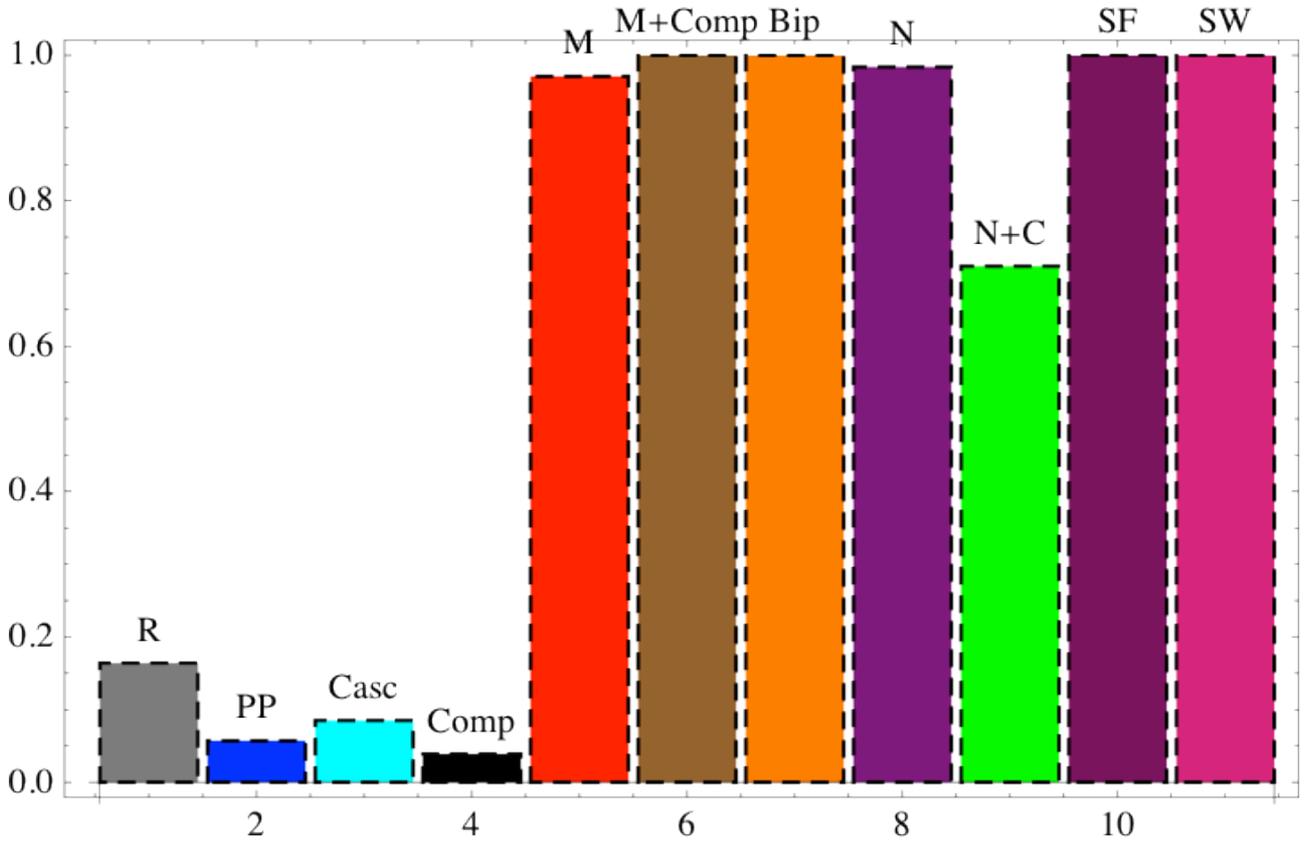

**Figure 4.** Statistics of the early signs detection. We first calculate the distribution of the correlation, $\rho_K$, between $Max[S_y]$ and the parameter $p$, after 1000 realizations for the full disordered (not mean-field) case. If $\rho_K$ is significant (p-value<0.05) and $\rho_K > 0.5$ the increase in $Max[S_y]$ is interpreted as an early warning sign. We calculate these detection statistics for several realizations of each network structure and determine the probability of detecting the early warning sign of instability. We consider eleven different network architectures typical of ecological or social networks, including random (R), predator-prey (PP), cascade (Casc), compartmentalized (Comp), mutualistic (M), bipartite (Bip), nested (N), nested with competition (N+C), scale free (SF), and small world (SW). These networks have different structures for the adjacency matrix and different combination of interaction types, i.e (++) mutualistic, (+-) antagonistic, (--) competitive or a combination of them (See Supplementary Information for more details).